
\newbox\hdbox%
\newcount\hdrows%
\newcount\multispancount%
\newcount\ncase%
\newcount\ncols
\newcount\nrows%
\newcount\nspan%
\newcount\ntemp%
\newdimen\hdsize%
\newdimen\newhdsize%
\newdimen\parasize%
\newdimen\spreadwidth%
\newdimen\thicksize%
\newdimen\thinsize%
\newdimen\tablewidth%
\newif\ifcentertables%
\newif\ifendsize%
\newif\iffirstrow%
\newif\iftableinfo%
\newtoks\dbt%
\newtoks\hdtks%
\newtoks\savetks%
\newtoks\tableLETtokens%
\newtoks\tabletokens%
\newtoks\widthspec%
\tableinfotrue%
\catcode`\@=11
\def\tstrut{\vrule height3.1ex depth1.2ex width0pt}%
\def\and{\char`\&}
\def\tablerule{\noalign{\hrule height\thinsize depth0pt}}%
\thicksize=1.5pt
\thinsize=0.6pt
\def\thickrule{\noalign{\hrule height\thicksize depth0pt}}%
\def\ctr#1{\hfil\ #1\hfil}%
%
%
%
%
\tablewidth=-\maxdimen%
\spreadwidth=-\maxdimen%
\def\tabskipglue{0pt plus 1fil minus 1fil}%
%
%
\centertablestrue%
%
%
%
%
\parasize=4in%
\gdef\ARGS{########}
\gdef\headerARGS{####}
\def\@mpersand{&}
{\catcode`\|=13
\gdef\letbarzero{\let|0}
\gdef\letbartab{\def|{&&}}%
\gdef\letvbbar{\let\vb|}%
}
{\catcode`\&=4
\def\ampskip{&\omit\hfil&}
\catcode`\&=13
\let&0
\xdef\letampskip{\def&{\ampskip}}%
\gdef\letnovbamp{\let\novb&\let\tab&}
}
\def\begintable{
   \begingroup%
   \catcode`\|=13\letbartab\letvbbar%
   \catcode`\&=13\letampskip\letnovbamp%
   \def\multispan##1{
      \omit \mscount##1%
      \multiply\mscount\tw@\advance\mscount\m@ne%
      \loop\ifnum\mscount>\@ne \sp@n\repeat%
   }
   \def\|{%
      &\omit\widevline&%
   }%
   \ruledtable
}
\long\def\ruledtable#1\endtable{%
   \offinterlineskip
   \tabskip 0pt
   \def\widevline{\vrule width\thicksize}
   \def\endrow{\@mpersand\omit\hfil\crnorm\@mpersand}%
   \def\crthick{\@mpersand\crnorm\thickrule\@mpersand}%
   \def\crthickneg##1{\@mpersand\crnorm\thickrule
          \noalign{{\skip0=##1\vskip-\skip0}}\@mpersand}%
   \def\crnorule{\@mpersand\crnorm\@mpersand}%
   \def\crnoruleneg##1{\@mpersand\crnorm
          \noalign{{\skip0=##1\vskip-\skip0}}\@mpersand}%
   \let\nr=\crnorule
   \def\endtable{\@mpersand\crnorm\thickrule}%
   \let\crnorm=\cr
   \edef\cr{\@mpersand\crnorm\tablerule\@mpersand}%
   \def\crneg##1{\@mpersand\crnorm\tablerule
          \noalign{{\skip0=##1\vskip-\skip0}}\@mpersand}%
   \let\ctneg=\crthickneg
   \let\nrneg=\crnoruleneg
   \the\tableLETtokens
   \tabletokens={&#1}
%
   \countROWS\tabletokens\into\nrows%
   \countCOLS\tabletokens\into\ncols%
   \advance\ncols by -1%
   \divide\ncols by 2%
   \advance\nrows by 1%
%
%
   \ifcentertables
      \ifhmode \par\fi
      \line{
      \hss
   \else %
      \hbox{%
   \fi
      \vbox{%
         \makePREAMBLE{\the\ncols}
         \edef\next{\preamble}
         \let\preamble=\next
         \makeTABLE{\preamble}{\tabletokens}
      }
      \ifcentertables \hss}\else }\fi
   \endgroup
   \tablewidth=-\maxdimen
   \spreadwidth=-\maxdimen
}
\def\makeTABLE#1#2{
   {
   \let\ifmath0
   \let\header0
   \let\multispan0
%
%
   \ncase=0%
   \ifdim\tablewidth>-\maxdimen \ncase=1\fi%
   \ifdim\spreadwidth>-\maxdimen \ncase=2\fi%
   \relax
%
   \ifcase\ncase %
      \widthspec={}%
   \or %
      \widthspec=\expandafter{\expandafter t\expandafter o%
                 \the\tablewidth}%
   \else %
      \widthspec=\expandafter{\expandafter s\expandafter p\expandafter r%
                 \expandafter e\expandafter a\expandafter d%
                 \the\spreadwidth}%
   \fi %
   \xdef\next{
      \halign\the\widthspec{%
      #1
      \noalign{\hrule height\thicksize depth0pt}
      \the#2\endtable
      }
   }
   }
   \next
}
\def\makePREAMBLE#1{
   \ncols=#1
   \begingroup
   \let\ARGS=0
   \edef\xtp{\widevline\ARGS\tabskip\tabskipglue%
   &\ctr{\ARGS}\tstrut}
   \advance\ncols by -1
   \loop
      \ifnum\ncols>0 %
      \advance\ncols by -1%
      \edef\xtp{\xtp&\vrule width\thinsize\ARGS&\ctr{\ARGS}}%
   \repeat
   \xdef\preamble{\xtp&\widevline\ARGS\tabskip0pt%
   \crnorm}
   \endgroup
}
\def\countROWS#1\into#2{
   \let\countREGISTER=#2%
   \countREGISTER=0%
   \expandafter\ROWcount\the#1\endcount%
}%
\def\ROWcount{%
   \afterassignment\subROWcount\let\next= %
}%
\def\subROWcount{%
   \ifx\next\endcount %
      \let\next=\relax%
   \else%
      \ncase=0%
      \ifx\next\cr %
         \global\advance\countREGISTER by 1%
         \ncase=0%
      \fi%
      \ifx\next\endrow %
         \global\advance\countREGISTER by 1%
         \ncase=0%
      \fi%
      \ifx\next\crthick %
         \global\advance\countREGISTER by 1%
         \ncase=0%
      \fi%
      \ifx\next\crnorule %
         \global\advance\countREGISTER by 1%
         \ncase=0%
      \fi%
      \ifx\next\crthickneg %
         \global\advance\countREGISTER by 1%
         \ncase=0%
      \fi%
      \ifx\next\crnoruleneg %
         \global\advance\countREGISTER by 1%
         \ncase=0%
      \fi%
      \ifx\next\crneg %
         \global\advance\countREGISTER by 1%
         \ncase=0%
      \fi%
      \ifx\next\header %
         \ncase=1%
      \fi%
      \relax%
      \ifcase\ncase %
         \let\next\ROWcount%
      \or %
         \let\next\argROWskip%
      \else %
      \fi%
   \fi%
   \next%
}
\def\counthdROWS#1\into#2{%
\dvr{10}%
   \let\countREGISTER=#2%
   \countREGISTER=0%
\dvr{11}%
\dvr{13}%
   \expandafter\hdROWcount\the#1\endcount%
\dvr{12}%
}%
\def\hdROWcount{%
   \afterassignment\subhdROWcount\let\next= %
}%
\def\subhdROWcount{%
   \ifx\next\endcount %
      \let\next=\relax%
   \else%
      \ncase=0%
      \ifx\next\cr %
         \global\advance\countREGISTER by 1%
         \ncase=0%
      \fi%
      \ifx\next\endrow %
         \global\advance\countREGISTER by 1%
         \ncase=0%
      \fi%
      \ifx\next\crthick %
         \global\advance\countREGISTER by 1%
         \ncase=0%
      \fi%
      \ifx\next\crnorule %
         \global\advance\countREGISTER by 1%
         \ncase=0%
      \fi%
      \ifx\next\header %
         \ncase=1%
      \fi%
\relax%
      \ifcase\ncase %
         \let\next\hdROWcount%
      \or%
         \let\next\arghdROWskip%
      \else %
      \fi%
   \fi%
   \next%
}%
{\catcode`\|=13\letbartab
\gdef\countCOLS#1\into#2{%
   \let\countREGISTER=#2%
   \global\countREGISTER=0%
   \global\multispancount=0%
   \global\firstrowtrue
   \expandafter\COLcount\the#1\endcount%
   \global\advance\countREGISTER by 3%
   \global\advance\countREGISTER by -\multispancount
}%
\gdef\COLcount{%
   \afterassignment\subCOLcount\let\next= %
}%
{\catcode`\&=13%
\gdef\subCOLcount{%
   \ifx\next\endcount %
      \let\next=\relax%
   \else%
      \ncase=0%
      \iffirstrow
         \ifx\next& %
            \global\advance\countREGISTER by 2%
            \ncase=0%
         \fi%
         \ifx\next\span %
            \global\advance\countREGISTER by 1%
            \ncase=0%
         \fi%
         \ifx\next| %
            \global\advance\countREGISTER by 2%
            \ncase=0%
         \fi
         \ifx\next\|
            \global\advance\countREGISTER by 2%
            \ncase=0%
         \fi
         \ifx\next\multispan
            \ncase=1%
            \global\advance\multispancount by 1%
         \fi
         \ifx\next\header
            \ncase=2%
         \fi
         \ifx\next\cr       \global\firstrowfalse \fi
         \ifx\next\endrow   \global\firstrowfalse \fi
         \ifx\next\crthick  \global\firstrowfalse \fi
         \ifx\next\crnorule \global\firstrowfalse \fi
         \ifx\next\crnoruleneg \global\firstrowfalse \fi
         \ifx\next\crthickneg  \global\firstrowfalse \fi
         \ifx\next\crneg       \global\firstrowfalse \fi
      \fi
\relax
      \ifcase\ncase %
         \let\next\COLcount%
      \or %
         \let\next\spancount%
      \or %
         \let\next\argCOLskip%
      \else %
      \fi %
   \fi%
   \next%
}%
\gdef\argROWskip#1{%
   \let\next\ROWcount \next%
}
\gdef\arghdROWskip#1{%
   \let\next\ROWcount \next%
}
\gdef\argCOLskip#1{%
   \let\next\COLcount \next%
}
}
}
\def\spancount#1{
   \nspan=#1\multiply\nspan by 2\advance\nspan by -1%
   \global\advance \countREGISTER by \nspan
   \let\next\COLcount \next}%
\def\dvr#1{\relax}%
\def\header#1{%
\dvr{1}{\let\cr=\@mpersand%
\hdtks={#1}%
\counthdROWS\hdtks\into\hdrows%
\advance\hdrows by 1%
\ifnum\hdrows=0 \hdrows=1 \fi%
\dvr{5}\makehdPREAMBLE{\the\hdrows}%
\dvr{6}\getHDdimen{#1}%
{\parindent=0pt\hsize=\hdsize{\let\ifmath0%
\xdef\next{\valign{\headerpreamble #1\crnorm}}}\dvr{7}\next\dvr{8}%
}%
}\dvr{2}}
\def\makehdPREAMBLE#1{
\dvr{3}%
\hdrows=#1
{
\let\headerARGS=0%
\let\cr=\crnorm%
\edef\xtp{\vfil\hfil\hbox{\headerARGS}\hfil\vfil}%
\advance\hdrows by -1
\loop
\ifnum\hdrows>0%
\advance\hdrows by -1%
\edef\xtp{\xtp&\vfil\hfil\hbox{\headerARGS}\hfil\vfil}%
\repeat%
\xdef\headerpreamble{\xtp\crcr}%
}
\dvr{4}}
\def\getHDdimen#1{%
\hdsize=0pt%
\getsize#1\cr\end\cr%
}
\def\getsize#1\cr{%
\endsizefalse\savetks={#1}%
\expandafter\lookend\the\savetks\cr%
\relax \ifendsize \let\next\relax \else%
\setbox\hdbox=\hbox{#1}\newhdsize=1.0\wd\hdbox%
\ifdim\newhdsize>\hdsize \hdsize=\newhdsize \fi%
\let\next\getsize \fi%
\next%
}%
\def\lookend{\afterassignment\sublookend\let\looknext= }%
\def\sublookend{\relax%
\ifx\looknext\cr %
\let\looknext\relax \else %
   \relax
   \ifx\looknext\end \global\endsizetrue \fi%
   \let\looknext=\lookend%
    \fi \looknext%
}%
%
%
\def\tablelet#1{%
   \tableLETtokens=\expandafter{\the\tableLETtokens #1}%
}%
\catcode`\@=12

\def\ifundefined#1{\expandafter\ifx\csname
#1\endcsname\relax}

\newcount\eqnumber \eqnumber=0
\def\beq{ \global\advance\eqnumber by 1 $$ }
\def\eeq{ \eqno(\the\eqnumber)$$ }
\def\label#1{\ifundefined{#1}
\expandafter\xdef\csname #1\endcsname{\the\eqnumber}
\else\message{label #1 already in use}\fi}
\def\(#1){(\csname #1\endcsname)}
\def\puteqno{\global\advance \eqnumber by 1 (\the\eqnumber)}

\newcount\refno \refno=0
\def\[#1]{\ifundefined{#1}\advance\refno by 1
\expandafter\xdef\csname #1\endcsname{\the\refno}
\fi[\csname #1\endcsname]}
\def\refis[#1]{\item{\csname #1\endcsname.}}


\baselineskip=18pt
\magnification=1200
\def\dim{{\rm dim}}
\def\ddt{{d\over{dt}}}
UR-1296

ER-745-40685

hep-th/9212021

\centerline{{\bf MAGNETIC FIELDS AND PASSIVE SCALARS IN}}
\centerline{{\bf POLYAKOV'S CONFORMAL TURBULENCE}}
\vskip .5in
\centerline{ G. Ferretti\footnote\dag{ BITNET address FERRETTI@UORHEP }}
\centerline{and}
\centerline{ Z. Yang\footnote\ddag{ BITNET address YANG@UORHEP }}
\centerline{\sl Department of Physics and Astronomy }
\centerline{\sl University of Rochester }
\centerline{\sl Rochester, NY 14627 }
\vskip .5in
\centerline{\bf Abstract}

Polyakov has suggested that two dimensional turbulence might
be described by a minimal model of conformal field theory.
However, there are many minimal models satisfying
the same physical inputs as Polyakov's
solution ${\cal M}_{(2,21)}$.
Dynamical magnetic fields and passive scalars pose different
physical requirements. For large magnetic Reynolds
number other minimal models arise.
The simplest one, ${\cal M}_{(2,13)}$,
makes reasonable predictions that may
be tested in the astrophysical context.
In particular, the equipartition theorem between
magnetic and kinetic energies does not hold:
the magnetic one dominates at larger distances.

\vfill\eject

Turbulence is one of the longstanding problems of theoretical physics.
It affects an enormous variety of phenomena, from galactic plasma to
cigarette smoke \[books]. As such,
the problem is far from being solved.
The randomness of the basic physical quantities clearly suggests that
the theory of random fields should be used to study this problem.
However, since one is dealing with a strongly coupled field
theory problem, the lack of reliable computational
techniques has slowed down the progress.

Recently, Polyakov \[polyakov] has made a bold attempt to understand
fully developed, homogeneous, isotropic, stationary two dimensional
turbulence by using conformal field theory.
Fully developed turbulence simply means that the motion of the fluid is
not affected by external constraints such as boundaries (at large
distances), or viscosity (at short distances). It is in this case
that universal features of turbulence have a better chance to be
understood. For fully developed turbulence, the assumptions of spatial
homogeneity (translation invariance) and isotropy (rotation invariance)
are very reasonable. More subtle is the stationary  assumption
(time translation invariance) because it implies that turbulence must be
sustained by external forces. The restriction to the two dimensional case
is, of course, because one wants to use the full power of the
conformal field theory \[belavin].

Polyakov's basic idea is to describe the universal features of two
dimensional turbulence by fitting it into a (non unitary)
minimal model ${\cal M}_{(p,q)}$ of two
dimensional conformal field theory,
here $1<p<q$, $p$ and $q$ integers, prime to each other \[belavin].
Any such model contains $(p-1)(q-1)/2$ primary fields $\Phi_{(m,n)}$,
$0<m<p$, $0<n<q$, $\Phi_{(m,n)}\equiv\Phi_{(p-m,q-n)}$
with conformal weight
\beq
    \Delta_{(m,n)}={{(pn-qm)^2-(p-q)^2}\over{4pq}}.
\eeq
For completeness, let us recall that the fusion rules are
\beq
    [\Phi_{(m_1,n_1)}]\times[\Phi_{(m_2,n_2)}]=
    \sum_{k=|m_1-m_2|+1}^{{\min(m_1+m_2-1,}\atop{2p-m_1-m_2-1)}}
    \hbox{\kern-.25in $'$ \kern.2in}
    \sum_{l=|n_1-n_2|+1}^{{\min(n_1+n_2-1,}\atop{2q-n_1-n_2-1)}}
    \hbox{\kern-.25in $'$ \kern.2in}
    [\Phi_{(k,l)}],
\eeq
where the symbol $\Sigma{}'$ means that the sum is performed with index
jumping of two by two, e.g. $k=3,5,7\cdots$.
The infrared contributions from the regular terms must be always kept in
mind because otherwise exact conformal symmetry would imply the
orthogonality theorem between primary fields.

In two dimensions, the velocity field $v_\alpha$ of an incompressible
fluid ($\partial_\alpha v_\alpha=0$), can be written in terms of a single
(pseudo) scalar function $\psi$, called the stream function:
\beq
    v_\alpha=\epsilon_{\alpha\beta}\partial_\beta\psi.
\eeq
The vorticity $\omega$ associated to the velocity field is a scalar
and can also be written purely in terms on the stream function:
\beq
    \omega=\epsilon_{\alpha\beta}\partial_\alpha v_\beta=
    -\partial_\alpha\partial_\alpha\psi. \label{om}
\eeq
One then attempts to interpret $\psi$ as one of the primary fields of some
minimal model $\psi=\Phi_{(m,n)}$
and to use the dynamics of the theory (Navier-Stokes
equations plus energy consideration) to fix $p$, $q$, $m$ and $n$.
Let us briefly recall how this is done in \[polyakov].

The Navier-Stokes equations for this two dimensional
incompressible fluid can be written, (after taking the curl) as
\beq
    \ddt\omega\equiv\dot\omega+\epsilon_{\alpha\beta}
    \partial_\beta\psi\partial_\alpha\omega=
    \nu\partial_\alpha\partial_\alpha\psi. \label{NS}
\eeq
The notation is as follows:
throughout this paper, the symbol $\ddt$ will denote the convective
derivative of any quantity, i.e.,
the time derivative along the motion of a
particle in the fluid. The ordinary partial time derivative will be
denoted by a dot or by $\partial_t$. The relation between the two is
\beq
    \ddt=\partial_t+v_\alpha\partial_\alpha\equiv\partial_t+
    \epsilon_{\alpha\beta}\partial_\beta\psi\partial_\alpha.
\eeq
The constant $\nu$ represents the kinematic viscosity. In the following we
will always work in the so called inertial range, where the effective
Reynolds number is huge and the viscous term in \(NS) can be neglected;
\beq
     \ddt\omega=0. \label{inviscidNS}
\eeq
One then talks of \(inviscidNS) as the inviscid Navier-Stokes equations.

In two dimensions, in addition to the kinetic energy per unit mass density
$E={1\over 2}\int v_\alpha v_\alpha d^2x$, there is another quadratic
conserved quantity. This new quantity is defined as the integral of the
square vorticity per unit mass density:
$H={1\over 2}\int \omega^2 d^2x$ and it is
known as the ``enstrophy''. From the detailed analysis in
\[kraichnan] it turns out that, for scales smaller than the scale at
which energy is fed into the system, the flux of enstrophy $\eta$
is the only relevant dimensional parameter and one could estimate the scaling
behavior of all observables in the Kolmogorov approximation.
For example, the energy spectrum $E(k)$, defined as
\beq
    {1\over 2}<v_\alpha v_\alpha> = \int d^1k E(k),
    \quad k=\sqrt{k_\alpha k_\alpha},
\eeq
is given, by dimensional analysis, as $E(k)\approx \eta^{2/3}k^{-3}$.
Of course, the aim is to find corrections to this scaling behavior.
{}From the definition of $E(k)$
and being $v_\alpha$ a level
one spin one field in $[\psi]$ one has
\beq
    E(k)\approx k^{-1+2(\Delta_v+\bar\Delta_v)}=k^{1+4\Delta_\psi},
\eeq
and what is left is to find $\Delta_\psi$.

In order to achieve this goal, one first translates
the constant flux of enstrophy into the scaling behavior
of the two point correlation function:
\beq
    <\dot\omega(r)\omega(0)>\approx r^0. \label{omegadotomega}
\eeq
Now, if the primary field $\psi$ satisfies the fusion rules
\beq
    [\psi]\times [\psi]=[\phi]+\hbox{less singular terms},
\eeq
and $\phi$ is neither the identity nor a level two degenerate field,
{}from \(inviscidNS) one has $\dot\omega\in[\phi]$
at level two and from \(om) one has
$\omega\in[\psi]$ at level one. Hence:
\beq
    \Delta_{\omega}=\bar\Delta_{\omega}=\Delta_{\psi}+1
    \quad\hbox{and}\quad
    \Delta_{\dot\omega}=\bar\Delta_{\dot\omega}=\Delta_{\phi}+2.
\eeq
Equation \(omegadotomega) then can be used to fix the sum of the conformal
weights:
\beq
    \Delta_{\dot\omega}+\bar\Delta_{\dot\omega}+
    \Delta_{\omega}+\bar\Delta_{\omega}=
    2(\Delta_{\phi}+2+\Delta_{\psi}+1)=0.
\eeq

The physical conditions coming from the inviscid Navier-Stokes equations
and the enstrophy flux then prompt us to look for a minimal model
${\cal M}_{(p,q)}$
containing two primary fields satisfying the following two
conditions:
\beq
    [\psi]\times [\psi]=[\phi]+\hbox{less singular terms;}
    \quad\hbox{and}\quad
    \Delta_{\phi}+\Delta_{\psi}=-3.
\eeq
Polyakov has found the following solution: ${\cal M}_{(2,21)}$, with
$\psi=\Phi_{(1,4)}$ and $\phi=\Phi_{(1,7)}$. Unfortunately, there are many
(perhaps infinitely many) other solutions to these conditions.
A very detailed list of solutions with energy spectrum steeper
than $k^{-3}$ is given very recently in \[matsuo]. We also
independently found the first few of them
with the help of a simple computer program.
In the table below we summarize the first few solutions we have found,
with no particular constraints on the energy spectrum.
The first one is Polyakov's solution. Together with his solution,
${\cal M}_{(3,25)}$, ${\cal M}_{(3,26)}$ and ${\cal M}_{(6,55)}$
give the most reasonable energy spectrum, i.e., steeper than
$k^{-3}$. These are among those found
by Matsuo in \[matsuo], where the condition on the steepness of the
spectrum is enforced by the extra constraint $\Delta_\psi<-1$.
Notice that only the first and
the last solutions are parity invariant, in the sense
$[\psi]\not\in[\psi]\times[\psi]$, but there are others,
not included in the table.
\bigskip
\centerline{First few solutions of Polyakov's requirements}
\bigskip
\midinsert
\thicksize=.5pt
\thinsize=.5pt
\tablewidth=5.5in
\begintable
$(p,q)$ |$\psi$         |$\phi$         |$E(k)\approx$\cr
$(2,21)$|$\Phi_{(1,4)}$ |$\Phi_{(1,7)}$ |$k^{-25/7}$  \cr
$(3,25)$|$\Phi_{(1,11)}$|$\Phi_{(1,9)}$ |$k^{-23/5}$  \cr
$(3,26)$|$\Phi_{(1,5)}$ |$\Phi_{(1,9)}$ |$k^{-55/13}$ \cr
$(3,35)$|$\Phi_{(1,21)}$|$\Phi_{(1,11)}$|$k^{-9/7}$   \cr
$(5,51)$|$\Phi_{(1,17)}$|$\Phi_{(1,11)}$|$k^{-47/17}$ \cr
$(6,55)$|$\Phi_{(1,14)}$|$\Phi_{(1,9)}$ |$k^{-41/11}$
\endtable
\vskip.1in
\endinsert\noindent

These results show that the requirements above are not
restrictive enough to pin down uniquely one minimal model.
Nevertheless, Polyakov's solution is somewhat special in
that it contains the smallest number of primary fields and
hence might be more stable. It is interesting to
include magnetic fields into the picture to see if
constraints are more stringent. But before we discuss this attempt,
let us first consider a simpler situation: the diffusion
of a passive scalar into Polyakov's ${\cal M}_{(2,21)}$ solution.

The term ``passive scalar'' denotes any scalar field such as temperature,
density of a pollutant etc..., that does not modify the Navier-Stokes
equations and satisfies the diffusion equation
\beq
    \ddt T = \kappa\partial_\alpha\partial_\alpha T. \label{passive}
\eeq
The assumption that $T$ does not appear into the Navier-Stokes equations
always implies some approximation. For instance, if $T$ stands for the
absolute temperature, the assumption is that the dilation coefficient of
the fluid is small, i.e.,
the density remains constant over the temperature
range being examined. From now on we will identify T with the absolute
temperature.

We are interested in the so called inertial-convective range,
where the Navier-Stokes equation is \(inviscidNS) and the equation for the
temperature reduces to $\ddt T=0$. The temperature spectrum is defined,
in analogy with the energy spectrum, as
\beq
    {1\over 2}<T^2>=\int d^1k E_T(k).
\eeq

Let us first look at the spectrum in the Kolmogorov approximation.
Let us define the temperature dissipation rate as $\epsilon_T$.
Clearly,
\beq
    \dim(E_T)=\dim (T^2)\times{\rm length}
    \quad\hbox{and}\quad
    \dim(\epsilon_T)=\dim (T^2)\times{\rm time}^{-1}
\eeq
The temperature spectrum must then be proportional to
$\epsilon_T$ and, since
in the enstrophy dominated inertial range the only other dimensional
parameter is $\eta$, one obtains the well known scaling law
\beq
    E_T(k)\approx \epsilon_T\eta^{-1/3}k^{-1}
\eeq
This is the Kolmogorov-type prediction of the spectrum in the
inertial-convective range. Note, in particular, that it is obeyed by
the vorticity $\omega$ itself when treated as a passive scalar, being
$E_\omega(k)=k^2 E(k)$ and $\epsilon_\omega=\eta$.

The universality of this behavior suggests that, in this regime,
the exponent for the temperature spectrum should be the same
as that for the vorticity spectrum, even if the latter is not simply
given by the Kolmogorov exponent $-1$ above. The ${\cal M}_{(2,21)}$
solution predicts $E_T(k)\approx k^{-11/7}$. In any case, it
is not correct to try to identify another primary field in
${\cal M}_{(2,21)}$ as the temperature because the predicted exponents
would be very far off.

Let us now turn to the main focus of this paper: two
dimensional turbulence in presence of magnetic fields.
In three dimensional space, the magnetohydrodynamics (MHD) equations
can be written, in the limit of infinitely conducting, incompressible
and inviscid fluid, purely as functions of the velocity
$\vec v$ and the magnetic field $\vec B$.
It is convenient to rescale the magnetic field as given in S.I. units by
the constant factor $1/\sqrt{\mu\rho}$, so that the rescaled field has
the dimensions of velocity. With this convention, taking the curl of the
Navier-Stokes equation to get rid of the pressure term,
the equations take the form
\beq\eqalign{
    \vec\nabla\cdot\vec v&=0\cr
    \vec\nabla\cdot\vec B&=0\cr
    \vec\nabla\times\bigg(\partial_t\vec v\bigg)&=\vec\nabla\times
    \bigg(-(\vec v\cdot\vec\nabla)\vec v+
    (\vec B\cdot\vec\nabla)\vec B\bigg)\cr
    \partial_t\vec B&=(\vec B\cdot\vec\nabla)\vec v-
    (\vec v\cdot\vec\nabla)\vec B\cr}\label{threedim}
\eeq

Let us now consider the case of an ``effectively two dimensional'' theory,
defined by the condition:
\beq
     \partial_3\vec v = 0\quad\hbox{and}\quad \partial_3\vec B = 0.
\eeq
With this assumption, one can treat the third components of the magnetic
field and of the velocity as a two dimensional scalar: $B_3\equiv B$ and
$v_3\equiv v$. Furthermore, one can introduce
two stream functions $A$ and $\psi$ for the
first two components of $\vec B$ and $\vec v$ respectively:
\beq
    B_\alpha=\epsilon_{\alpha\beta}\partial_\beta A
    \quad\hbox{and}\quad
    v_\alpha=\epsilon_{\alpha\beta}\partial_\beta\psi.
\eeq
For this effectively two dimensional theory, equations \(threedim) can
be written simply in terms of four independent two dimensional scalars:
$A$, $B$, $v$ and $\psi$. These equations look particularly nice if one
also introduces (as dependent quantities) the two dimensional curls
\beq
    J\equiv\epsilon_{\alpha\beta}\partial_\alpha B_\beta=
    -\partial_\alpha\partial_\alpha A
    \quad\hbox{and}\quad
    \omega\equiv\epsilon_{\alpha\beta}\partial_\alpha v_\beta=
    -\partial_\alpha\partial_\alpha \psi,
\eeq
and the differential operator:
\beq
    {\cal A}=\epsilon_{\alpha\beta}\partial_\beta A\partial_\alpha.
\eeq
With the aid of the quantities defined above, the equations of motion can
than be written as
\beq\eqalign{
    \ddt \omega&={\cal A} J\cr
    \ddt v&={\cal A} B\cr
    \ddt B&={\cal A} v\cr
    \ddt A&=0.\cr}\label{twodim}
\eeq
This shows that the fields $B$ and $v$ only appear in the second and
the third equations of \(twodim). It is therefore consistent to study first
the possible turbulent solution of $A$ and $\psi$ by setting $B$
and $v$ to zero.

Before we go any further, one important remark must be made, in order to
avoid confusion. The kind of turbulent solution we are considering is not
the same as the kind of two dimensional turbulence that has recently been
studied in laboratory \[sommeria].
In laboratory experiments, one is dealing with magnetic Reynolds numbers
that are fairly small and the system (typically a tank filled with
mercury) is subjected to a strong external magnetic field
perpendicular to the plane. One then
writes the total magnetic field as the sum of the external one
$\vec B_{{\rm ext.}}$ and the turbulent one $\vec b$. Since the turbulent
component is much smaller than the external field, the MHD equations can be
expanded in first order in $\vec b$.
After taking the curl, the magnetic field drops out
completely of the Navier-Stokes equation and its only
effect is to reduce the $\vec B=0$ three dimensional
Navier-Stokes equation to the $\vec B=0$ two dimensional Navier-Stokes
equation \(NS). One could then study the behavior of $\vec b$ as a passive
vector on the ordinary turbulent background. As far as conformal field
theory goes, one is then brought back to Polyakov's solution.

In the following, we consider a different regime altogether.
We assume that there
is no external magnetic field and that the turbulent magnetic field is
parallel to the plane and is
(in our units) comparable with the velocity. This can happen only at very
high magnetic Reynolds number, such as in astrophysical situations
(for instance in a galaxy).
One last word of caution: we are not interested in the dynamo effect.
As a matter of fact, the dynamo effect does not take place at all
in two dimensions \[zildovich]. In two dimensions,
in order to keep turbulence in the
stationary regime, one has to pump in magnetic energy as well as kinetic
energy, otherwise the magnetic field would dissipate due to the finite
conductivity of the material just as much as the velocity would dissipate
due to the finite viscosity.

The third component of the magnetic field need
not be ignored but can be treated at a
later stage since it decouples from
the equations for $\psi$ and $A$ as already stressed.
In any case, one cannot invoke this magnetic field as the agent that
blocks the system into the two dimensional state. However, if one is
willing to take the astrophysical scenario seriously, there are other
possible effects that might help, such as the rotation of the galaxy or
gravity itself.

Let us start by rewriting the equations for $\psi$ and $A$ more
explicitly
\beq\eqalign{
 \dot\omega+\epsilon_{\alpha\beta}\partial_\beta\psi\partial_\alpha\omega
 &=\epsilon_{\alpha\beta}\partial_\beta A\partial_\alpha J\cr
 \dot A +\epsilon_{\alpha\beta}\partial_\beta\psi\partial_\alpha A&=0.\cr}
 \label{explicit}
\eeq
{}From \(explicit) we see that a radical
change takes place in this regime.
Enstrophy is no longer a conserved quantity. Its place is taken by the
integral of the square of the stream function $A$. The only two quadratic
conserved quantities are then
\beq
    E={1\over 2}\int\;( v_\alpha v_\alpha+B_\alpha B_\alpha) d^2x
    \quad\hbox{and}\quad
    G={1\over 2}\int A^2 d^2x.
\eeq
It is then natural to look for a solution for which
\beq
    <\dot A(r)A(0)>\approx r^0. \label{adota}
\eeq
Instead of the constant enstrophy condition, we are considering
$\epsilon_A$, the dissipation rate of $A$, as a constant.

Let us translate these physical conditions in the language of conformal
field theory. Again, the working assumption is that we try to fit the
requirements \(explicit) and \(adota) by a minimal model.
We interpret the two basic scalars $A$ and $\psi$ as primary fields
and $\dot A$ as a secondary field in the conformal family
$[\chi]$ of the most singular field
appearing in the fusion rules
of $A$ and $\psi$. By using point split regularization,
one can see that $\dot A$ must be a level two scalar in $[\chi]$.
\vfill\eject

We must then look for a minimal model ${\cal M}_{(p,q)}$
satisfying the following requirements:
\item{a)} There are two primary fields $A$ and $\chi$ for which
$\Delta_A+\Delta_\chi=-2$.
\item{b)} There is another primary field $\psi$ such that its fusion rules
are, up to less singular terms, $[\psi]\times [A]=[\chi]$.

To be sure, these conditions do not fix the theory uniquely. The simplest
solution however, provides a nice example. Surprisingly, this solution is
simpler (i.e., contains less primary fields) than Polyakov's. It is the
minimal model ${\cal M}_{(2,13)}$, whose basic structure is summarized
in the table below.
\bigskip
\centerline{Structure and interpretation of ${\cal M}_{(2,13)}$}
\bigskip
\midinsert
\thicksize=.5pt
\thinsize=.5pt
\tablewidth=5.5in
\begintable
|$\Phi_{(1,1)}$|$\Phi_{(1,2)}$|$\Phi_{(1,3)}$|$\Phi_{(1,4)}$|
$\Phi_{(1,5)}$|$\Phi_{(1,6)}$\cr
meaning:|$I$|$\psi$|?|$A$|$\chi$|?\cr
$\Delta=$|$0$|$-5/13$|$-9/13$|$-12/13$|$-14/13$|$-15/13$
\endtable
\vskip.1in
\endinsert\noindent
It should not be surprising that this solution is not the same as the
one before. Remember that the introduction
of magnetic fields adds an extra
dimensional parameter to the theory. This parameter is still relevant in
the inertial range and it affects the scaling laws already in
Kolmogorov theory.

Having found a solution for our physical requirements, it is now a simple
exercise in conformal field theory to work out its basic predictions.
The kinetic energy and magnetic energy spectra behave respectively as
\beq
    E_{{\rm kin.}}(k)\approx k^{-7/13}
    \quad\hbox{and}\quad
    E_{{\rm mag.}}(k)\approx k^{-35/13}.
\eeq
This means that the equipartition theorem does not hold
exactly for our solution.
In fact, the kinetic energy will dominate at short distance and the
magnetic one will take over at large distance.

Passive scalars will still be represented by level one fields.
But now, the argument given above using the vorticity no
longer apply, because $\omega$ no longer satisfies the equation
$\ddt \omega=0$. For example, if one takes $T\in[\Phi_{(1,6)}]$,
level one scalar, one finds $E_T(k)\approx k^{-21/13}$ but
we do not see any physical reason for that identification.

In conclusion, Polyakov's idea gives a fresh new look on one of the oldest
unsolved problems of physics. Unfortunately, the physical requirements found
so far are not enough to uniquely identify one specific minimal model.
In fact, there are perhaps infinitely many minimal models satisfying the
constraints. One wonders whether the ``correct'' conformal field theory
might be the one obtained as some limiting case of the minimal
models above or perhaps as some other, non minimal, conformal field theory.
Quite apart from this, the introduction of magnetic fields and passive
scalars is of some interest. It is still not enough to fix the theory
completely but the simplest model satisfying the physical requirements gives
sensible predictions that might be relevant in the astrophysical context.

We would like to tank S.G. Rajeev for his interest and his helpful
comments. We also benefitted from discussions with A. Das and P. Bedaque.
This work is supported in part by DOE Grant DE-FG02-91ER40685.
\vfill\eject

\centerline{\bf References}

\refis[books] L.D. Landau and E.M. Lifshitz, {\it Course of
              Theoretical Physics} \hfill\break
              Pergamon Press, Vol. 6 and 8.
              \hfill\break
              M. Lesieur, {\it Turbulence in Fluids}, Kluwer Academic
              Publishers
              \hfill\break
              R. Moreau, {\it Magnetohydrodynamics}, Kluwer Academic
              Publishers

\refis[polyakov] A.Polyakov, PUPT-1341, hep-th/9209046
                 Preprint (1992).

\refis[belavin] A.A. Belavin, A. Polyakov and A.B. Zamolodchikov,
                \hfill\break Nucl. Phys. {\bf B241} (1984) 333.

\refis[kraichnan] R.H. Kraichnan, Phys. of Fluids {\bf 10} (1967) 1417.

\refis[matsuo] Y Matsuo, UT-620, hep-th/9212010 Preprint (1992).

\refis[sommeria] J. Somm\'eria, J. Fluid Mech. {\bf 170} (1986) 139.

\refis[zildovich] I.B. Zil'dovich, Soviet Physics JETP {\bf 4}
                  (1957) 460.
\bye